# Astrometric and photometric observations of comet 29P/Schwassmann--Wachmann 1 at the Sanglokh international astronomical observatory

G.I. Kokhirova [a,*], O.V. Ivanova [b,c,d], F. Dzh. Rakhmatullaeva [a], A.M. Buriev [a], U.Kh. Khamroev [a]

[a] *Institute of Astrophysics of the Academy of Sciences of the Republic of Tajikistan, Bukhoro 22, Dushanbe, 734042, Tajikistan*
[b] *Astronomical Institute of the Slovak Academy of Sciences, SK-05960 Tatranska Lomnica, Slovak Republic*
[c] *Main Astronomical Observatory of the National Academy of Sciences, 27 Akademika Zabolotnoho, Street, 03143, Kyiv, Ukraine*
[d] *Astronomical Observatory, Taras Shevchenko National University of Kyiv, 3 Observatorna St., 04053, Kyiv, Ukraine*

A B S T R A C T

Astrometric and photometric observations of the comet 29P/Schwassmann-Wachmann 1 were performed at the Zeiss-1000 telescope of the International Astronomical Observatory Sanglokh (IAOS) of the Institute of Astro-physics, Academy of Sciences of the Republic of Tajikistan in July–August 2017. Although the comet has a short period of revolution it is regarded to be an object of the Centaurs group. Comet was exhibited a new activity this period which we used for analysis of its features. The coordinates of comet were determined and the orbit was calculated, the apparent and absolute magnitudes in BVRI bands were determined, as well the comet color indices and the estimation of nucleus diameter were obtained. From investigations of morphological features we iden-tified two dust structures in the coma.

## Introduction

A new short-period comet was discovered at the Hamburg Observatory in Germany on 15 November 1927. In honor of the discoverers, it was named comet 29P/Schwassmann-Wachmann 1, hereinafter on text 29P. The revolution period of 29P around the Sun is 14.6 years and it has been observed since then. Alternatively, for 29P the Tisserand parameter relatively Jupiter is $T_j = 2.984$ (CNEOS, 2019). These facts allowed classifying 29P as a short-period comet. However, then a new class of small bodies was identified, called the Centaurs, which includes objects having both the perihelion distance $q$ and the semi-major axis $a$ of their orbits located between the orbits of Jupiter (at the heliocentric distance of 5.2 au) and Neptune (at the distance of 30 au). These objects are also characterized by chaotic orbits (Bailey and Malhotra, 2009). It turned out that the indicated parameters of the 29P orbit satisfy this condition, and therefore 29P was assigned to the group of Centaurs (Jewitt and Kalas, 1998). The orbital elements of 29P are arranged in 1, where $a$ is the semi-major axis, $e$ is the eccentricity, $q$, $Q$ are the perihelion and aphelion distances, $i$ is the inclination, ω is the argument of perihelion, Ω is the longitude of perihelion (CNEOS, 2019).

The duality of 29P is caused by a very small eccentricity value, as a consequence of which it moves around the Sun in almost circular orbit unusual for comets. To present more than 45 such objects are known, being ice bodies and locating between the orbits of Jupiter and Neptune. The orbit of 29P is outside the Jupiter orbit located at the heliocentric distance of 5 au. Note, in the Solar system this distance is considered to be the formal boundary, so-called "snow line", starting from which the temperature becomes low enough so that the solid phase of water and other volatile compounds is stable even under the solar radiation.

It is assumed that the Centaurs are objects transferred from the Kuiper belt to the inner region of the planetary system, where they are appearing as short-period comets during ground-based observations (Jewitt and Luu, 1993>; Jewitt and Kalas, 1998; Levison and Duncan, 1997; Jewitt et al., 1998). So, scientifically, the Centaurs are interesting as nearest ones, i.e. brighter and more accessible samples of the Kuiper belt objects.

The physical properties of 29P found from ground-based observations also point to its duality. It is known that the geometric albedo value usually varies from 0.02 to 0.12 for cometary nuclei, the average value is 0.07 (Jewitt, 1991). The geometric albedo of 29P found from photometric measurements in the visible spectral range is $p_V = 0.13$ (Cruikshank and Brown, 1983), that is completely atypical for the cometary nucleus. However, this albedo value is typical for objects of the Centaur group (Barucci et al., 2004). There are various estimations of the albedo value of 29P ranging from 0.02 to 0.17 that were measured in different radiation ranges. In particular, for the geometric albedo of cometary nucleus a value of 0.033 adopted (CNEOS, 2019).

**Table 1**
Orbital elements of 29P (J2000.0)

| Epoch | $a$ (au) | $e$ | $q$ (au) | $Q$ (au) | $i$ (deg.) | ω (deg.) | Ω (deg.) |
|---|---|---|---|---|---|---|---|
| January 19, 2010 | 5.990 | 0.045 | 5.720 | 6.260 | 9.391 | 49.049 | 312.632 |

A large scatter in the estimates of the effective size of 29P nucleus is associated with this dispersion in the geometric albedo values, because an albedo is used to determine the radius. Radius estimates found using the photometric data by ground-based observations of 29P and assuming the geometric albedo of 0.04 lie in the range from 21 to 52 km (Cruikshank and Brown, 1983; Lamy et al., 2004; Meech et al., 1993). For the rotation period of 29P there are also several estimates from 10 h (Luu and Jewitt, 1993) to 14 h (Meech et al., 1993). By the observations of 29P from 2008 to 2009 the rotation period of 11.7–12.1 h was obtained (Ivanova et al., 2012). The total absolute magnitude of 29P is $M1 = 6.0^m$ and diameter is $d = 60.4$ km (CNEOS, 2019).

Since a discovery, 29P has become known for numerous outbursts, i.e. a sudden strong increase in brightness when the comet's magnitude in-creases by $2–5^m$ (Whipple, 1980; Sekanina, 1982; Wyckoff, 1982; Ivanova et al., 2009, 2016). As a result of a continuous monitoring during 2002–2007 28 outbursts of 29P (an average of 7.3 outbursts per year) were registered, moreover, it was shown that there is no clear periodicity in the appearance of outbursts, which confirms the unpredictability of the activity of this comet (Trigo-Rodriguez et al., 2008). However, Miles et al. (2016) found some periodicity in the outbursts in long term monitoring of 29P. Outbursts of short-period comets are usually associated with a splitting of their nuclei, but this process is once and not long (Boehnhardt, 2004). Outbursts can also occur when comets pass perihelion near the Sun. However, the orbit of 29P locates quite far from the Sun so that the surface temperature of its nucleus is certainly below the sublimation temperature of water ice. Therefore, other physical processes very likely should cause the appearance of 29P outbursts.

According to Froeschle et al. (1983) outbursts of brightness may be associated with a crystallization of amorphous ice on the nucleus surface. Later this suggestion was strongly supported by the detection of $CO$ gas emissions due to its release from amorphous ice (Senay and Jewitt, 1994). Additionally it was shown by Trigo-Rodriguez et al. (2008) that the outbursts are due to an increase of the gas-producing activity of 29P, and first of all, an increase in the generation of neutral $CO$ gas. While as shown by Ivanova et al. (2016) the amount of gases released during the exothermic phase of a transition of water ices from amorphous to the crystalline state, is insufficient to form the observed outburst activity of 29P. Therefore, the reliable reasons for the activity of 29P are still not fully defined.

Only a small number of comets from the Centaurs group have been studied, so it is very important to develop a complex investigation of such objects. Since a source of the Centaurs is the Kuiper belt, it is of particular interest to study the composition of the ices of their nuclei and its comparison with the composition of the nuclei of long-period comets originating from the Oort cloud. As already noted the mechanisms responsible for the occurrence of cometary activity at remote heliocentric distances have not yet been established.

## Observations, data processing and results

29P once again showed activity in 2017 and became available for observation. The astrometric and photometric observations of 29P were carried out at the Zeiss-1000 telescope of the International Astronomical Observatory Sanglokh (IAOS) of the Institute of Astrophysics, Academy of Sciences, Republic of Tajikistan on 28 July - 1 August 2017. The telescope is equipped with a FLI Proline PL16803 CCD camera; the focal distance of the telescope (the Cassegrainian focus) is $F = 13.3$ m, and the scale of the image is 63 μm/arcsec. The sensor is arranged as a nominally 4096 × 4096 array and the field of view is 11' × 11', implying in an image scale of 0.16" per pixel. To reduce the information redundancy and improve the signal-to-noise ratio $S/N$, the images were stored with using a binning value of 2, which made the working scale of the images to be 0.36" per pixel. The image quality, which was measured as an average of the full width at half maximum (FWHM) for several stars from individual images, was at a level of 2.1". We used the standard BVRI filters that rather closely reproduce the filters of the Johnson-Cousins photometric system. To reduce the noise level the CCD matrix was cooled to a temperature of −20°C. To take into account the dark current, align the flat fields of the images, and take into account the CCD camera errors, we used, respectively, "Dark", "Flat", and "Bias" exposures which were also used in the image processing. During observations more than 550 images with exposure time of 60 s were obtained from which 400 best images were used for reduction. The dates and time of observations of 29P (expressed in the UT day fractions), the number of acquired images $N$, the exposure time $t$, as well as the geocentric $r$ and heliocentric Δ distances, the position angle of direction to the Sun $PA$, and the phase angle $ph$ of 29P are given in Table 2. The image of 29P is presented in Fig. 1. To measure the images of 29P and the field stars, we used a fixed radius aperture, which allowed the object to be completely covered.

**Table 2**
Summary of observations of 29P at the Sanglokh observatory.

| Date, UT | $r$ (au) | Δ (au) | $PA$ (deg.) | $ph$ (deg.) | $N\times$ Band | $t$ (s) |
|---|---|---|---|---|---|---|
| 28.84, 2017 | 5.829 | 4.842 | 67.6 | 2.583 | $10 \times V$, $20 \times R$, $10 \times I$ | 60 |
| 29.88, 2017 | 5.828 | 4.838 | 67.2 | 2.403 | $8 \times V$, $11 \times R$, $11 \times I$ | 60 |
| 30.77, 2017 | 5.828 | 4.834 | 66.8 | 2.223 | $30 \times B$, $33 \times V$, $30 \times R$, $30 \times I$ | 60 |
| 31.85, 2017 | 5.828 | 4.831 | 66.2 | 2.042 | $40 \times B$, $40 \times V$, $40 \times R$, $40 \times I$ | 60 |
| 01.76, 2017 | 5.828 | 4.828 | 65.7 | 1.860 | $20 \times B$, $20 \times V$, $20 \times R$, $20 \times I$ | 60 |

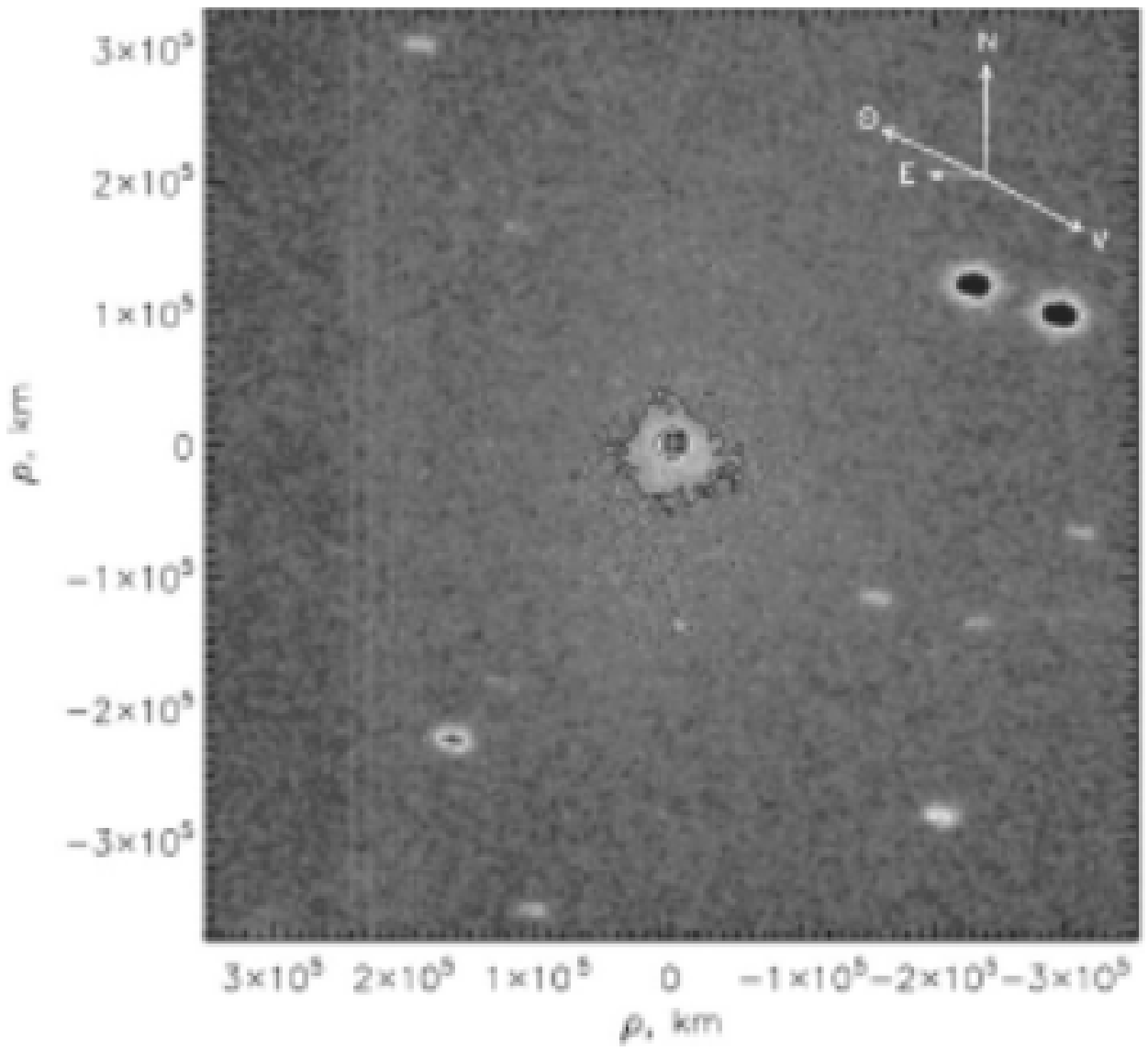

**Fig. 1.** Resultant image of 29P in R band obtained at Zeiss-1000 telescope on July 30, 2017, exposure 1800 s.

Since the Centaurs are characterized by chaotic motion it is important to periodically examine their orbits. Additionally, any significant changes in the orbital elements found during observations might be caused by a collision with some body. Such impact will be certainly responsible for the outbursts of an object and thus can be considered as one of possible mechanism that generated these outbursts. To determine the orbital parameters of 29P the astrometric measurements were included into observations. The astrometric reduction of the IAOS observations was performed using the APEKS-II software package developed at the Pulkovo Observatory (Devyatkin et al., 2010). The package calibrates the exposures, distinguishes the images of the stars and the objects, and identifies the stars according to the specified catalogs. In the astrometric reduction process, the image distortions produced by the optical system are taken into account with the six- and eight-constant methods (depending on a number of the identified stars). The astrometric UCAC5 catalog was used as a reference. The coordinates of stars with a magnitude between $10^m$ and $14^m$ are given in the UCAC5 with an astrometric accuracy of approximately 0.02", while the accuracy in the position of fainter stars (the limit is $16^m$) is roughly 0.07". To measure the positions in the frames, the reference stars with the brightness corresponding to the specified interval were used. The astrometric-reduction error averaged over all exposures is 0.04" and 0.09" for the right ascension α and the declination δ, respectively. The mean deviations of the measured equatorial coordinates *O* from the catalog data *C* designated as $(O-C)_\alpha$ and $(O-C)_\delta$ for the coordinates α and δ, respectively, and their corresponding mean-square errors $\sigma_\alpha$ and $\sigma_\delta$ are presented in Table 3. The results of determining the equatorial coordinates of 29P according to the astrometric reduction of IAOS observations are shown in Fig. 2, where the right ascension α and the declination δ of 29P are plotted on the abscissa and the ordinate, respectively (Kokhirova et al., 2017).

**Table 3**
Precision of observations of 29P at the Sanglokh observatory.

| Date | $(O-C)_\alpha^{''}$ | $\sigma_\alpha^{''}$ | $(O-C)_\delta^{''}$ | $\sigma_\delta^{''}$ |
|---|---|---|---|---|
| 28.84, 2017 | −0.023 | 0.018 | 0.073 | 0.038 |
| 29.88, 2017 | −0.062 | 0.026 | 0.071 | 0.015 |
| 30.77, 2017 | −0.063 | 0.025 | 0.086 | 0.019 |
| 31.85, 2017 | −0.080 | 0.044 | 0.093 | 0.034 |
| 01.76, 2017 | −0.056 | 0.016 | 0.035 | 0.019 |

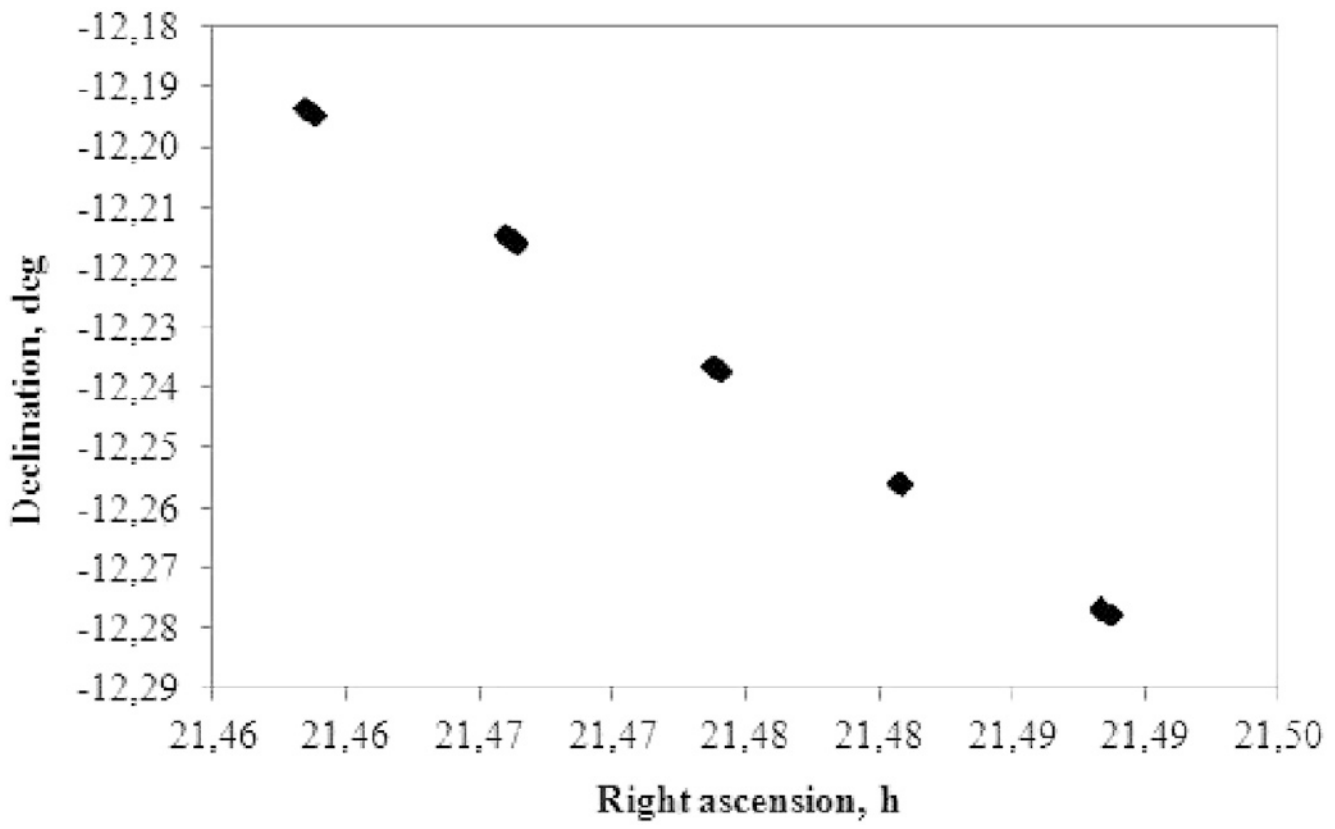

**Fig. 2.** The apparent trajectory of 29P derived from observation data at the Sanglokh observatory during July 28-August 1, 2017.

**Table 4**
Comparison of the initial orbit of 29P obtained from the Sanglokh observations to the MPC orbit (J2000.0).

| Orbital elements | SIAO (this work) | MPEC 2016-V116 | Δ |
|---|---|---|---|
| Number of positions used | | | |
| for orbit calculation | 300 | 20553 | – |
| *T* (JD) | 2458564.1256 | 2458567.2912 | – |
| Epoch (JD) | 2457820.5 | 2457600.5 | – |
| *e* | 0.0417716 | 0.0416442 | 0.0001274 |
| *a* (AU) | 6.01622181 | 6.0151515 | 0.00107031 |
| *q* (AU) | 5.76491423 | 5.7646555 | 0.00025873 |
| *i* (deg.) | 9.37444 | 9.37658 | −0.00214 |
| ω (deg.) | 48.75893 | 48.98332 | −0.22439 |
| Ω (deg.) | 312.40993 | 312.40927 | 0.00066 |
| *n* (deg./day) | 0.06679105 | 0.06680885 | −0.00000178 |
| σ (arcsec) | 0.288 | 0.600 | – |

The orbit of 29P can be constructed based on its measured coordinates. First, from several observations, the initial orbit is constructed; and it is defined more precisely later, with the use of additional new observations. The orbit of comet was determined with the EPOS software package developed also at the Pulkovo Observatory (L'vov & Tsekmeister, 2012). For the mean time moment of 300 observations performed at the IAOS the initial orbit of comet calculated and presented in Table 4 where, for comparison, an orbit is also given, calculated from observations available in the IAU Minor planet center (Williams, 2016). In Table 4 alongside with usual orbital elements the following data are also given: the moment of perihelion passage, the Epoch of orbit's calculation, the mean-diurnal motion *n*, the period of revolution *P*, and the mean-square errors σ; the last column contains the corrections. As seen, a quite recognizable comet orbit was obtained by the observations at the Sanglokh observatory. Analysis of the results of astrometry shows sufficient accuracy in determining the coordinates from observations at the Sanglokh observatory, and the orbit elements calculated from obtained coordinates pointed to this. It may be concluded that the orbit of 29P at the time of observation is stable within the obtained accuracy.

The photometric reduction was performed using adopted standard procedure. To make absolute photometric measurements, all of the field stars, which had been preliminarily examined for variability, were considered. The catalog of the Photometric All-Sky Survey of the American Association of Variable Star Observers (abbreviated as APASS) (Henden et al., 2011) was used for photometric studies. The APASS catalog contains the stars with magnitudes in the interval between $7^m$ and $17^m$ measured in five transmission bands: B and V of the Johnson - Cousins system and *g*', *r*', and *i*' of the Sloan system. To pass from the Sloan system to the Johnson - Cousins one, the transition equations for the R band were taken from the paper by Mallama (2014). To process the photometric images in a standard way, we made master-frames of the zero exposure and dark and flat fields. All the frames containing the comet's images were corrected for zero-point and nonuniformity of the pixel sensitivity with the use of the master-frames. The sky background was determined with the standard IDL procedure called Sky (Landsman, 1993). For the aperture photometry of stars, the diaphragm with a radius of 3" (3 × FWHM) was used. The residual sky background was estimated with a circular aperture. To calculate the stellar magnitude error, we summed up the statistical errors, which are caused by the $S/N$ ratio for the object and the reference stars, and the errors in the catalog magnitudes of standard stars (for the APASS catalog, the error for the standard stars was assumed as $0.03^m$ (Henden et al., 2011)).

The light curves of 29P obtained in such a way are shown in Fig. 3, where the apparent magnitudes *m* in BVRI bands and the observation dates (expressed in Julian days) are on the ordinate and the abscissa, respectively. The mean values of the apparent magnitudes of the object in different bands for each observational night are presented in Table 5 where the apparent brightness is expressed in stellar magnitudes estimated within the aperture radius (the aperture size is 3", which corresponds to ~10518 km).

The apparent magnitude $m_a$ was converted to the absolute brightness $m_a(1,1,0)$ of the comet nucleus according to the following empirical formula (Snodgrass et al., 2006)

$$m_a(1,1,0) = m_a - 5\log(r\Delta) - \beta(ph), \quad (1)$$

where $m_a(1,1,0)$ (or *H*) is the brightness of hypothetic point at unitary heliocentric and geocentric distances with the phase angle $ph = 0$ deg., $m_a$ is the measured magnitude, *r* and Δ are the helio- and geocentric distances of comet in au, *ph* is the phase angle in degree, β is the phase coefficient in mag/deg.; subscript $a$ indicates the band. The value of phase coefficient $\beta = 0.035$ mag/deg. was used (Lamy et al., 2004). The absolute brightness of 29P in the BVRI bands (the averages for one night) found in such a way are listed in Table 6, the light curves are presented in Fig. 4. As seen, the absolute magnitude of comet was practically permanent during monitoring.

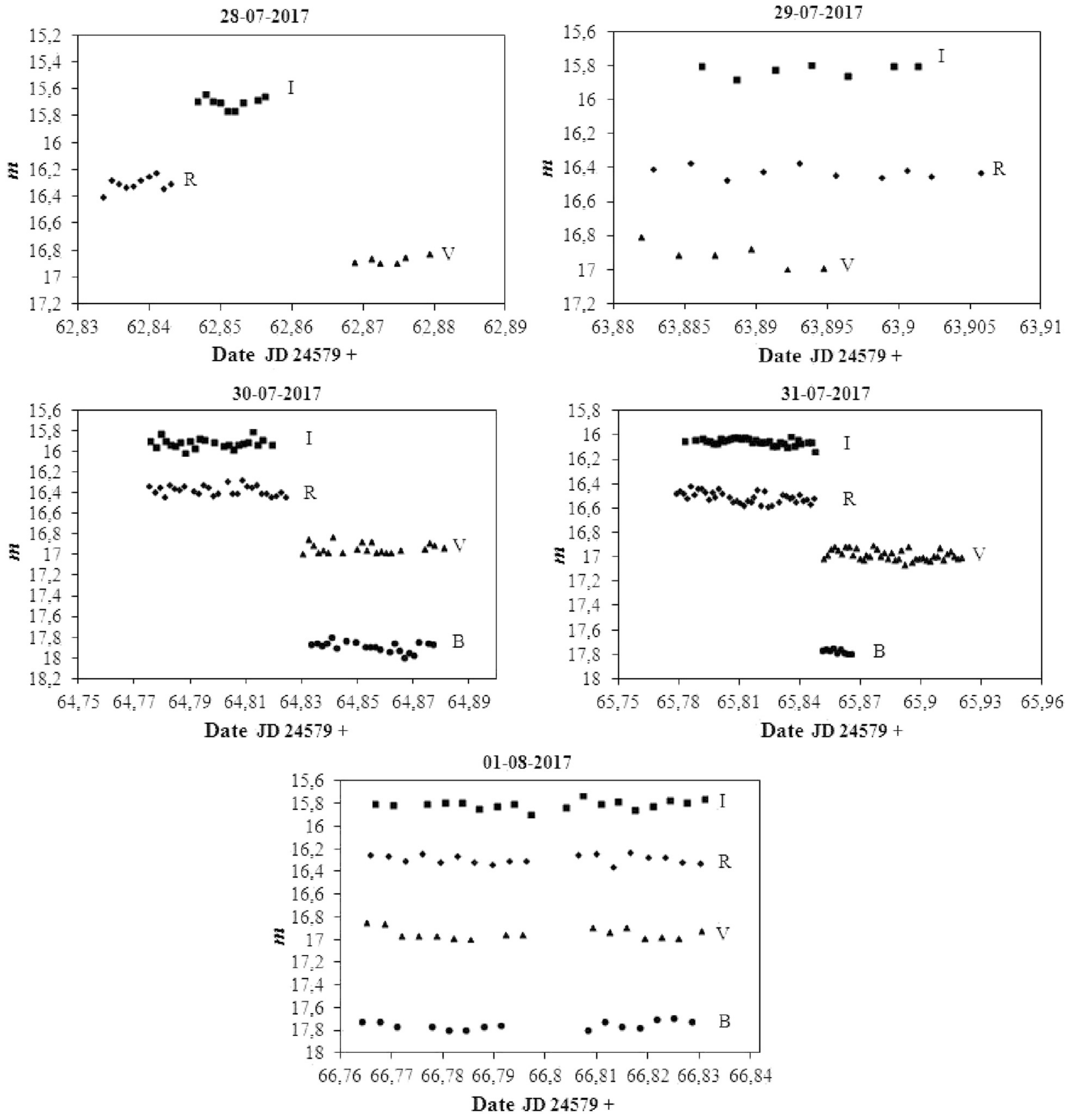

**Fig. 3.** The light curves of 29P in the BVRI bands according to the Sanglokh observations on 28 July-1 August 2017.

**Table 5**
The apparent brightness of 29P according to the Sanglokh observations in 2017.

| Bands | July 28, 2017 | July 29, 2017 | July 30, 2017 | July 31, 2017 | August 01, 2017 |
|---|---|---|---|---|---|
| *B* | – | – | 17.78±0.05 | 17.77±0.02 | 17.75±0.04 |
| *V* | 16.88±0.04 | 16.91±0.05 | 16.92±0.05 | 16.97±0.04 | 16.92±0.04 |
| *R* | 16.29±0.04 | 16.41±0.04 | 16.45±0.05 | 16.50±0.04 | 16.33±0.04 |
| *I* | 15.69±0.03 | 15.83±0.05 | 15.92±0.05 | 16.05±0.02 | 15.82±0.04 |

The mean values of the color indices of 29P according to our observations as well as of other objects of the Solar System, like active Jupiter family comets [JFC] (Solontoi et al., 2012), active long-period comets [LPC] (Jewitt, 2015), Kuiper belt objects [KBO] (Tegler, 2015), active and inactive Centaurs [Centaurs] (Jewitt, 2015), and the Sun [Sun] (Holmberg et al., 2006) are listed in Table 7. As seen, the color indices of 29P determined by the IAOS monitoring are relevant to the mean color indices of the active objects of the Centaurs group and Kuiper belt objects. Color indices indicate a shift of the radiation maximum to the red part of the spectrum, which implies the dominant contribution of dust particles into the formation of coma. According to observations of the cometary activity during the period 2002–2007 it was also concluded that coma of 29P is continuously replenished with fine dust released from the surface of the nucleus (Trigo-Rodriguez et al., 2008). The systematic nature of this process is also confirmed by observations of the Spitzer space telescope which recorded dust jets of 29P while a lack of comet outbursts (Stansberry et al., 2004).

**Table 6**
The absolute brightness of 29P $m_a(1,1,0)$ in the BVRI bands according to the Sanglokh observations.

| Bands | July 28, 2017 | July 29, 2017 | July 30, 2017 | July 31, 2017 | August 01, 2017 |
|---|---|---|---|---|---|
| *B* | – | – | 10.45±0.05 | 10.45±0.02 | 10.45±0.02 |
| *V* | 9.54± 0.04 | 9.57±0.05 | 9.60±0.05 | 9.65±0.03 | 9.61±0.04 |
| *R* | 8.95± 0.04 | 9.07±0.04 | 9.13±0.04 | 9.18±0.04 | 9.01±0.02 |
| *I* | 8.19± 0.03 | 8.50±0.04 | 8.59±0.04 | 8.73±0.02 | 8.51±0.04 |

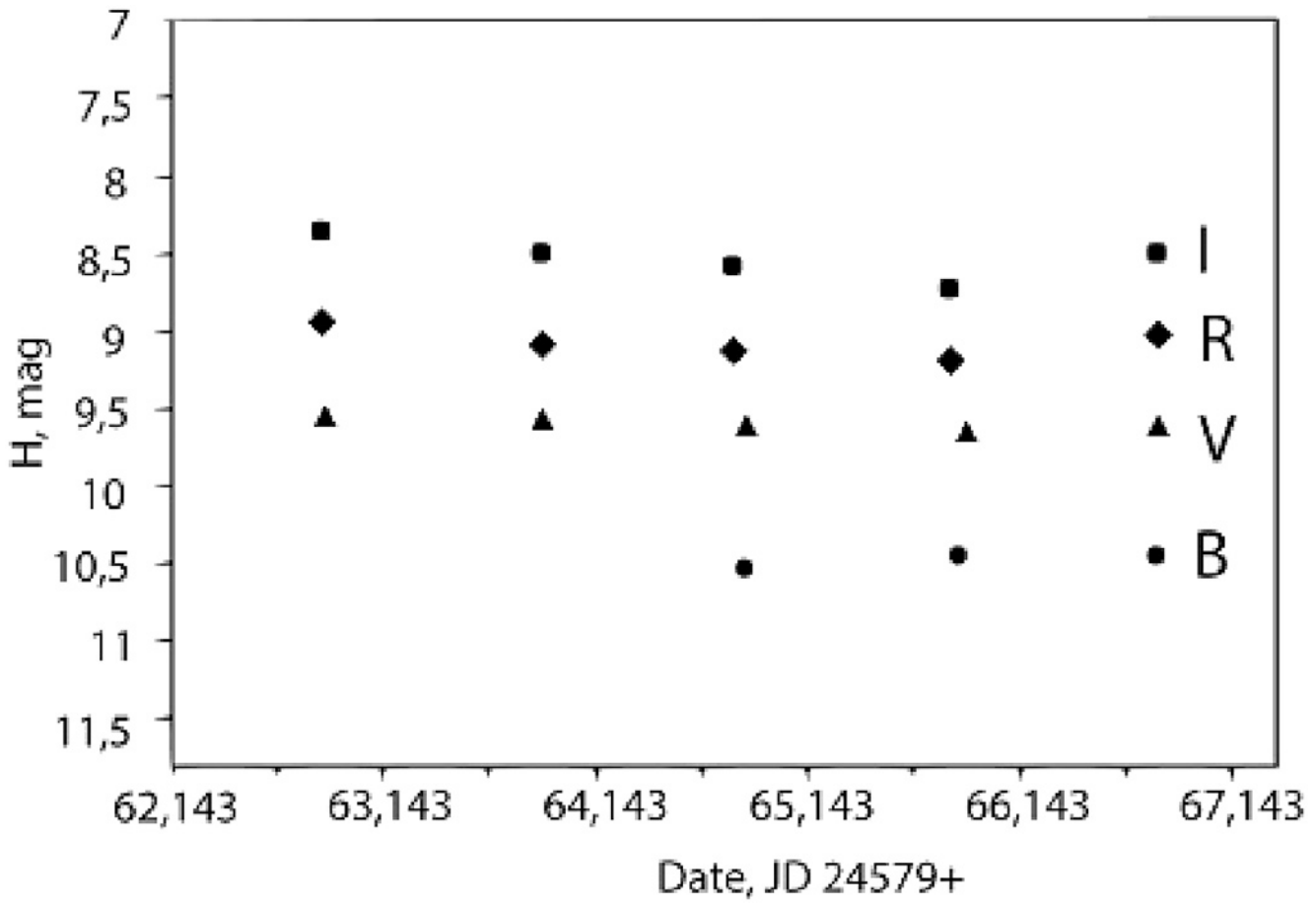

**Fig. 4.** The absolute magnitude *H* (mean values per night) of 29P in BVRI bands by the Sanglokh observations during 28 July-1 August 2017.

To estimate the size of 29P nucleus, we used the following empirical expression between the apparent magnitude measured in the V band $m_V$ and the effective radius of cometary nucleus $r_N$ (expressed in meters) (Russell, 1916)

$$Ar_N^2 = 2.238 \cdot 10^{22} R_h^2 \Delta^2 \cdot 10^{0.4(m_\odot - m_V + \beta\alpha)}, \qquad (2)$$

where $R_h$ is heliocentric distance in au. Once the absolute magnitude in the V band $m_V(1,1,0)$ was found then the radius can be calculated by a simplified version of equation (2):

$$Ar_N^2 = 2.238 \cdot 10^{22} \cdot 10^{0.4(m_\odot - m_V(1,1,0))}, \qquad (3)$$

where $A$ is the geometric albedo and $m_\odot$ = -26.75 is the apparent magnitude of the Sun, both values are given in the V band. Since the exact albedo value of 29P is not known we adopt the interval 0.13> $A$ >0.03 in our calculations for the nucleus size. The estimates of the cometary nu-cleus diameter according to our absolute brightness measurements $m_V$ in the V filter are presented in Table 8. Note we presented the upper limit of effective diameter of 29P computed under the assumption of two values of the geometric albedo. Actually, the size estimation strongly depends on the geometric albedo and phase function which are unknown for 29P.

As it was noted, when the geometric albedo value is assumed to be 0.04 the estimate of the nucleus radius lies in the range from 21 to 52 km. According to Cruikshank and Brown (1983) the nucleus radius is estimated as 20 km when the geometric albedo value of 0.13. The estimates of diameter obtained by our measurements range from 43.3 to 45.7 km for $A = 0.13$, and 86.7–91.4 km when $A = 0.033$ and is quite consistent to available data. To clarify the albedo and the size of the nucleus, more observations of 29P are necessary.

*Morphology*

As was mentioned above, 29P presents outburst activity with big numbers of structures in cometary coma (Berman and Whipple, 1928a, b; Trigo-Rodriguez et al., 2010; Ivanova et al., 2016; Miles et al., 2016; Picazzio et al., 2019). For our data obtained with R and I filters we used an enhancement technique to segregate low-contrast structures in cometary coma. Before applying the filtering, images were cleaned from field stars around the cometary nucleus and coma. As a digital filter, we used the Larson-Sekanina algorithm (Larson and Sekanina, 1984). Fig. 5 shows direct and processed images using digital filter. We applied the digital filters to all individual exposures of the comet as well as to the same composite image to help in evaluating whether revealed features are real or not. As it is seen in the figures, the cometary coma is more condensed in R filter than I filter. The bright coma is 5000 km across, although an elongation of the coma is visible in the north direction. After processing of images with digital filter, 29P exhibits two dust structures, which can be seen across the images at sunward and anti-sunward directions.

**Table 7**
The mean color indices of 29P and other objects of the Solar System.

| index | 28.07 | 29.07 | 30.07 | 31.07 | 01.08 | JFC | LPC | KBOs | active | inact. | |
|---|---|---|---|---|---|---|---|---|---|---|---|
| Color | 29P in 2017 (this work) | | | | | Active | Active | Active | Centaurs | | Sun |
| *B* − *V* | – | – | 0.85 | 0.80 | 0.84 | 0.75 | 0.78 | 0.92 | 0.80 | 0.93 | 0.64 |
| *V* − *R* | 0.59 | 0.50 | 0.47 | 0.47 | 0.60 | 0.47 | 0.47 | 0.57 | 0.50 | 0.55 | 0.35 |
| *R* − *I* | 0.76 | 0.57 | 0.54 | 0.45 | 0.50 | 0.43 | 0.42 | – | 0.57 | 0.45 | 0.33 |
| *B* − *R* | – | – | 1.32 | 1.27 | 1.44 | 1.22 | 1.24 | 1.49 | 1.30 | 1.47 | 0.99 |

**Table 8**
Estimates of the size of the comet 29P nucleus by the IAOS observations.

| Date | $r$ (AU) | Δ(AU) | *ph* (deg.) | $m_V$ (mag.) | $m_V(1,1,0)$ (mag.) | $D$ (km) $A = 0.13$ | $D$ (km) $A = 0.033$ |
|---|---|---|---|---|---|---|---|
| 28.84, 2017 | 5.829 | 4.842 | 2.583 | 16.88±0.04 | 9.54±0.04 | 45.87±0.08 | 91.05±0.08 |
| 29.88, 2017 | 5.828 | 4.838 | 2.403 | 16.91±0.05 | 9.57±0.05 | 45.08±0.08 | 89.47±0.08 |
| 30.77, 2017 | 5.828 | 4.834 | 2.223 | 16.92±0.05 | 9.60±0.05 | 44.55±0.08 | 86.38±0.08 |
| 31.85, 2017 | 5.828 | 4.831 | 2.042 | 16.97±0.04 | 9.65±0.03 | 43.52±0.08 | 86.39±0.08 |
| 01.76, 2017 | 5.828 | 4.828 | 1.860 | 16.92±0.04 | 9.61±0.04 | 44.27±0.08 | 87.86±0.08 |

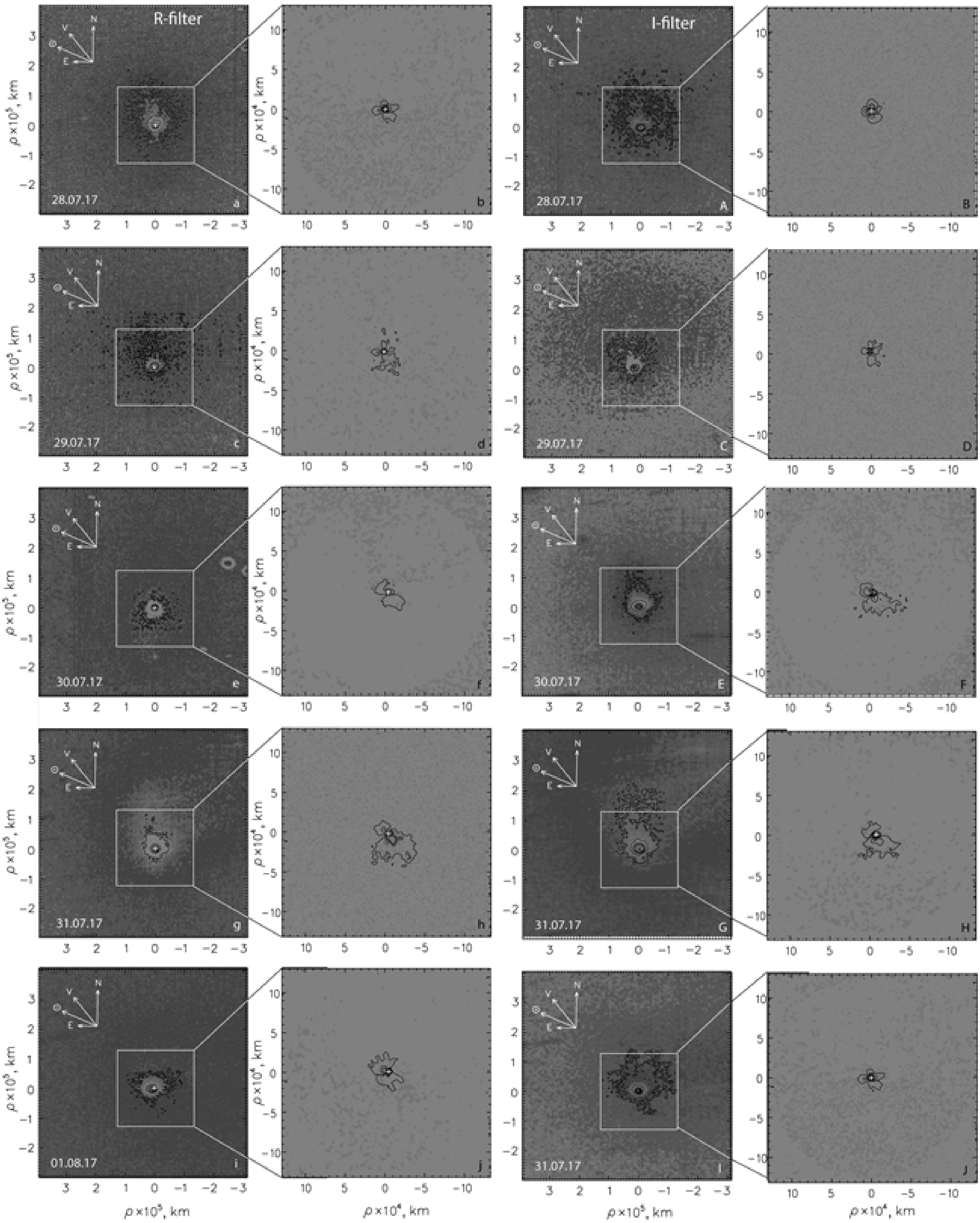

**Fig. 5.** Images of comet 29P in the R and I filters obtained from July 28 to August 1, 2017. Frames a, c, e, g, i and A, C, E, G, I show the direct images of 29P in relative intensity in R and I filters, respectively. Frames b, d, f, h, j and B, D, F, H, J represent intensity images to which was applied the Larson-Sekanina algorithm (Larson and Sekanina, 1984; Samarasinha and Larson, 2014). North (N), East (E), sunward (⊙), and velocity vector (*V*) directions are indicated on each direct frame.

## Conclusions

The astrometric and photometric observations of comet 29P carried out at the Sanglokh observatory in 2017 yielded the following results:

(1) The equatorial coordinates, geocentric trajectory and orbit of comet were determined.
(2) The apparent magnitudes of comet in BVRI filters were obtained and the light curves were plotted.
(3) The absolute brightness of comet in BVRI filters was determined. During the observational period the light curves showed no noticeable oscillations in the absolute brightness within the error measurements. The mean value of the absolute brightness of comet in the V and R filters was $9.60^{\mathrm{m}}\pm0.04^{m}$ and $9.10^{\mathrm{m}}\pm 0.04^{m}$, respectively.
(4) The color indices agree well with the currently available mean values for active objects of the Centaurs group asteroids. They point to the dominant contribution of the dust component into the coma which is redder in reflectance.
(5) The estimates of the nucleus diameter by our observations are matching to available published estimations.
(6) For our data obtained with R and I filters we used an enhancement technique to segregate low-contrast structures in cometary coma. It is shown that the cometary coma is more condensed in R filter than I filter; comet exhibits two dust structures at sunward and anti-sunward directions.

## Declaration of competing interest

The authors declare that they have no conflicts of interest.

## Acknowledgments

We would like to express our gratitude to the anonymous referees for discussion and useful comments that improved the paper.

OVI thanks the Slovak Academy of Sciences (grant Vega 2/0023/18).

## References

Bailey, B., Malhotra, R., 2009. Two dynamical classes of Centaurs. Icarus 203, 155–163.

Barucci, A.M., Doressoundiram, A., Cruikshank, D.P., 2004. Surface characteristics of transneptunian objects and Centaurs from photometry and spectroscopy. In: Festou, M.C., Keller, H.U., Weaver, H.A. (Eds.), Comets II. Univ. Arizona Press, Tucson, AZ, pp. 647–658.

Berman, L., Whipple, F.L., 1928a. Notes on comet J 1927. Publ. Astron. Soc. Pac. 40, 34.

Berman, L., Whipple, F.L., 1928b. Elements and ephemeris of comet J 1927 (Schwassmann-Wachmann). In: Lick Observ. Bull., vol. 394. Univ. of California Press, pp. 117–119.

Boehnhardt, H., 2004. Split comets. In: Festou, M.C., Keller, H.U., Weaver, H.A. (Eds.), Comets II. Univ. Arizona Press, Tucson, AZ, pp. 301–316.

CNEOS, 2019. Center for Near Earth Objects Studies accessed in 2019. https://cneos.jpl.nasa.gov/.

Cruikshank, D.P., Brown, R.H., 1983. The nucleus of comet P/Schwassmann-Wachmann 1. Icarus 56, 377–380.

Devyatkin, A.V., Gorshanov, D.L., Kouprianov, V.V., Verestchagina, I.A., 2010. Apex I and Apex II software packages for reduction of astronomical CCD observation. Sol. Syst. Res. 44, 68–80.

Froeschle, Cl, Klinger, J., Rickman, H., 1983. Thermal models for the nucleus of comet 29P/Schwassmann-Wachmann 1. In: Asteroids, Comets, Meteors; Proceed. Of the Meeting, Uppsala, Sweden, June 20-22, (A85-26851 11-89. Astronomiska Observatoriet, Uppsala, Sweden, pp. 215–224, 1983.

Holmberg, J., Flynn, C., Portinari, L., 2006. The colours of the Sun. MNRAS 367, 449–453.

Henden, Arne A., Levine, S.E., Terrell, D., Smith, T.C., Welch, D.L., 2011. Data release 3 of the AAVSO all-sky photometric Survey (APASS). In: American Astron. Soc., AAS Meeting, vol. 218. BAAS, p. 43 id.126.01.

Ivanova, A.V., Korsun, P.P., Afanasiev, V.L., 2009. Photometric investigations of distant comets C/2002 VQ94 (LINEAR) and 29P/Schwassmann-Wachmann-1. Sol. Syst. Res. 43, 453–462.

Ivanova, A.V., Afanasiev, V.L., Korsun, P.P., Baranskii, A.R., Andreev, M.V., Ponomarenko, V.A., 2012. The rotation period of comet 29P/Schwassmann-Wachmann 1 determined from the dust structures (Jets) in the coma. Sol. Syst. Res. 46, 313–319.

Ivanova, O.V., Luk'yanyk, I.V., Kiselev, N.N., Afanasiev, V.L., Picazzio, E., Cavichia, O., de Almeida, A.A., Andrievsky, S.M., 2016. Photometric and spectroscopic analysis of Comet 29P/Schwassmann-Wachmann 1 activity. Planet. Space Sci. 121, 10–17.

Jewitt, D., 1991. Cometary Photometry, vol. 116. Cambridge University Press, pp. 19–65, 1.

Jewitt, D., 2015. Color systematics of comets and related bodies. Astron. J. 150, 201–219.

Jewitt, D., Kalas, P., 1998. Thermal observations of Centaur 1997 CU26. Astrophys. J. 499, L103–L106.

Jewitt, D., Luu, J., Trujillo, C., 1998. Large Kuiper belt objects: the Mauna Kea 8K CCD Survey. Astron. J. 115, 2125–2135.

Jewitt, D., Luu, J., 1993. Discovery of the candidate Kuiper belt object 1992 QB1. Nature 362, 730–732.

Kokhirova, G.I., Ivanova, O.A., Buriev, A.M., Khamroev, U.Kh, Ibrohimov, A.A., Mullo-Abdolov, A.Sh, Rakhmatullaeva, F.Dzh, 2017. Observation of Comet C/2015 VL62 (Lemmon-Yeung-PANSTARRS), 29P/Schwassmann-Wachmann 1, Asteroids (596) Scheila, (190166) 2005 UP156 - MPC 105577, 5 October 2017.

Lamy, P.L., Toth, I., Fernandez, Y.R., Weaver, H.A., 2004. The sizes, shapes, albedos, and colors of cometary nuclei. In: Festou, M.C., Keller, H.U., Weaver, H.A. (Eds.), Comets II. Univ. Arizona Press, Tucson, AZ, pp. 223–264.

Landsman, W.B., 1993. The IDL astronomy user's library. In: Hanisch, R.J., Brissenden, R.J.V., Barnes, J. (Eds.), Astron. Data Anal. Software and Systems II, A.S.P. Conf. Ser., vol. 52, p. 246.

Larson, S.M., Sekanina, Z., 1984. Coma morphology and dust-emission pattern of periodic Comet Halley. I - high-resolution images taken at Mount Wilson in 1910. Astron. J. 89, 571–578.

Levison, H.F., Duncan, M.J., 1997. From the Kuiper belt to jupiter-family comets: the spatial distribution of ecliptic comets. Icarus 127 (1), 13–32.

L'vov, V.N., Tsekmeister, S.D., 2012. The use of the EPOS software package for research of the solar system objects. Sol. Syst. Res. 46, 177–179.

Luu, J.X., Jewitt, D., 1993. Periodic Comet Schwassmann- Wachmann 1. IAU Circ, p. 5692.

Mallama, A., 2014. Sloan magnitudes for the brightest stars. J. Am. Assoc. Var. Star Obs. 42, 443.

Meech, K.J., Belton, M.J.S., Mueller, B.E.A., Dicksion, M.W., Li, H.R., 1993. Nucleus properties of P/Schwassmann- Wachmann 1. Astron. J. 106, 1222–1236.

Miles, R., Faillace, G.A., Mottola, S., et al., 2016. Anatomy of outbursts and quiescent activity of Comet 29P/Schwassmann-Wachmann. Icarus 272, 327–355.

Picazzio, E., Luk'yanyk, I.V., Ivanova, O.V., Zubko, E., Cavichia, O., Videen, G., Andrievsky, S.M., 2019. Comet 29P/Schwassmann-Wachmann 1 dust environment from photometric observation at the SOAR Telescope. Icarus 319, 58–67.

Russell, H.N., 1916. On the albedo of the planets and their satellites. Astrophys. J. 43, 173–196.

Samarasinha, N.H., Larson, S.M., 2014. Image enhancement techniques for quantitative investigations of morphological features in cometary comae: a comparative study. Icarus 239, 168–185.

Sekanina, Z., 1982. The problem of split comets in review. In: Comets. (A83-13376 03-90) Univ. Arizona Press. University of Arizona Press, Tucson, AZ, pp. 251–287, 1982.

Senay, M.C., Jewitt, D., 1994. Coma formation driven by carbon monoxide release from comet Schwassmann-Wachmann 1. Nature 371, 229–231.

Snodgrass, C., Lowry, S.C., Fitzsimmons, A., 2006. Photometry of cometary nuclei: rotation rates, colours and a comparison with Kuiper Belt Objects. MNRAS 373, 1590–1602.

Stansberry, J.A., Cleve, V., Reach, W.T., Cruikshank, D.P., Emery, J.P., 2004. Spitzer observations of the dust coma and nucleus of 29P/Schwassmann-Wachmann 1. Astrophys. J. Suppl. 54, 463–468.

Solontoi, M., Ivezic, Z., Juric, M., Becker, A.C., et al., 2012. Ensemble properties of comets in the sloan digital sky Survey. Icarus 218, 571–584.

Tegler, S.C., 2015. Kuiper belt object magnitudes and surface colors. http://www.physics.nau.edu/tegler/research/survey.htm. (Accessed July 2019).

Trigo-Rodriguez, J.M., Melendo, E.C., Davidsson, B.J.R., Sanchez, A., Rodriguez, D., Lacruz, J., Reyes, A.D., Pastor, S., 2008. Outburst activity in comets I. Continuous monitoring of comet 29P/Schwassmann-Wachmann 1. Astron. Astrophys. 485, 509–606.

Trigo-Rodriguez, J.M., Garcia-Hernandez, D.A., Sanchez, A., et al., 2010. Outburst activity in comets - II. A multiband photometric monitoring of comet 29P/Schwassmann-Wachmann 1. MNRAS 409, 1682–1690.

Wyckoff, S., 1982. Overview of comet observations. In: Comets. (A83-13376 03-90) Univ. Arizona Press. Univ. Arizona Press, Tucson, pp. 3–55, 1982.

Whipple, F.L., 1980. Rotation and outbursts of comet P/Schwassmann-Wachmann 1. Astron. J. 85, 305–313.

Williams, G.V., 2016. 29P/Schwassmann-Wachmann 1. MPEC 2016-V116.